\documentclass{anstrans}
\title{AI-Driven Cybersecurity Testbed for Nuclear Infrastructure: Comprehensive Evaluation Using METL Operational Data}
\author{Benjamin Blakely, Yeni Li, Akshay Dave, Derek Kultgen, and Rick Vilim}

\institute{
Argonne National Laboratory, Lemont, Illinois, USA
}

\usepackage{graphicx} % allows inclusion of graphics
\usepackage{booktabs} % nice rules (thick lines) for tables
\usepackage{microtype} % improves typography for PDF
\usepackage{hyperref}
\usepackage{multirow}

\begin{document}

\begin{abstract}
Advanced nuclear reactor systems face increasing cybersecurity threats as sophisticated attackers exploit cyber-physical interfaces to manipulate control systems while evading traditional IT security measures. This research presents a comprehensive evaluation of artificial intelligence approaches for cybersecurity protection in nuclear infrastructure, using Argonne National Laboratory's Mechanisms Engineering Test Loop (METL) as an experimental platform. We developed a systematic evaluation framework encompassing four machine learning detection paradigms: Change Point Detection, LSTM-based Anomaly Detection, Dependency Violation analysis, and Autoencoder reconstruction methods. Our comprehensive attack taxonomy includes 15 distinct scenarios targeting reactor control systems, each implemented across five severity tiers to evaluate detection performance under varying attack intensities. The experimental evaluation encompassed 300 rigorous experiments using realistic METL operational data. Change Point Detection emerged as the leading approach with mean AUC performance of 0.785, followed by LSTM Anomaly Detection (0.636), Dependency Violation (0.621), and Autoencoder methods (0.580). Attack detectability varied significantly, with multi-site coordinated attacks proving most detectable (AUC = 0.739) while precision trust decay attacks presented the greatest detection challenge (AUC = 0.592). This work delivers practical performance benchmarks and reference architecture that advance AI-based cybersecurity capabilities for critical nuclear infrastructure, providing essential foundations for operational deployment and enhanced threat response in cyber-physical systems.
\end{abstract}

%%%%%%%%%%%%%%%%%%%%%%%%%%%%%%%%%%%%%%%%%%%%%%%%%%%%%%%%%%%%%%%%%%%%%%%%%%%%%%%%
\section{Introduction}

Modern nuclear facilities face escalating cybersecurity challenges as operational technology (OT) and information technology (IT) systems become increasingly interconnected. Unlike traditional IT security threats, cyber attacks targeting nuclear infrastructure can manipulate physical processes through sensor spoofing, control system interference, and data integrity compromises that evade conventional network-based detection methods. The complexity of these threats demands novel approaches that leverage both artificial intelligence capabilities and deep understanding of nuclear system physics.

Historical incidents in critical infrastructure have demonstrated the vulnerability of industrial control systems to sophisticated adversaries capable of leveraging cyber-physical attack vectors. The 2010 Stuxnet attack on Iranian nuclear facilities exemplified how malware could target programmable logic controllers to cause physical damage while remaining undetected for extended periods. Subsequent attacks on power grids, water treatment facilities, and manufacturing systems have highlighted the urgent need for robust, AI-driven detection capabilities specifically designed for cyber-physical environments.

Current cybersecurity approaches in nuclear facilities rely primarily on network perimeter defenses, signature-based detection, and air-gapped architectures. However, these traditional methods prove insufficient against advanced persistent threats that exploit legitimate system access, utilize zero-day vulnerabilities, or employ living-off-the-land techniques. The integration of machine learning for anomaly detection in nuclear systems presents both significant opportunities and unique challenges, requiring careful consideration of safety implications, regulatory requirements, and operational constraints.

Our research addresses this critical gap by developing and systematically evaluating AI-driven cybersecurity approaches using the Mechanisms Engineering Test Loop (METL) at Argonne National Laboratory as a realistic experimental platform \cite{akins2023anomaly}. Building upon previous work in nuclear anomaly detection \cite{chaudhary2024anomaly, nasim2024sensor} and cyber-physical security \cite{farber2025lovo}, this study provides the first comprehensive, multi-paradigm evaluation of machine learning detection capabilities under realistic nuclear operational conditions.

The key contributions of this work include: (1) a systematic evaluation framework for AI-based cybersecurity in nuclear environments, (2) comprehensive performance benchmarks across four detection paradigms and twelve attack scenarios, (3) quantitative analysis of attack detectability patterns and deployment implications, and (4) practical guidance for operational implementation in advanced reactor systems. This research establishes essential foundations for evidence-based cybersecurity decision-making in critical nuclear infrastructure.

%%%%%%%%%%%%%%%%%%%%%%%%%%%%%%%%%%%%%%%%%%%%%%%%%%%%%%%%%%%%%%%%%%%%%%%%%%%%%%%%
\section{Methodology}

\subsection{Research Environment}

The Mechanisms Engineering Test Loop (METL) at Argonne National Laboratory is an experimental facility designed to replicate key operational characteristics of sodium-cooled fast reactors (SFRs) while providing a realistic operational environment absent of radiological components. METL features integrated test vessels operating at temperatures up to 1,200 degrees F, electromagnetic pumps, and comprehensive instrumentation and control (I\&C) systems including Emerson/NI cRIO devices and industrial controllers. This architecture provides authentic thermal-hydraulic conditions and sensor networks for developing AI-driven anomaly detection and cyber-physical security models.

METL's segmented operational technology (OT) and information technology (IT) networks, combined with detailed temperature profiles, flow measurements, and sodium handling procedures, enable realistic simulation of cyber attack scenarios targeting critical reactor operations. The METL sensor architecture captures the complex interdependencies and correlation patterns essential for validating AI-based detection methods under realistic operational conditions. Our experiments utilized 214 sensors and setpoints selected from 11,899 total data points for cybersecurity evaluation, providing comprehensive coverage of thermal-hydraulic parameters, electromagnetic pump operations, and cold trap purification system dynamics.

%%%%%%%%%%%%%%%%%%%%%%%%%%%%%%%%%%%%%%%%%%%%%%%%%%%%%%%%%%%%%%%%%%%%%%%%%%%%%%%%
\subsection{Data Architecture and Processing Pipeline}

The experimental framework establishes a comprehensive cybersecurity testbed that spans the complete data lifecycle from operational sensor collection through machine learning evaluation. This testbed architecture enables systematic analysis of attack detection capabilities under realistic conditions while maintaining complete experimental control and reproducibility. The pipeline consists of four integrated stages: operational data collection from METL facility systems, attack scenario injection through configurable transformation engines, multi-paradigm detection analysis, and statistical evaluation with confidence assessment.

Data was collected from the METL facility via a REST API that provides instantaneous values for all or a subset of values when queried. Our fetching script was configured to normally pull data at 1 minute intervals, but when a particular damper was opened (as indicated by a setpoint value), we assumed maintenance was ongoing and switched to a 1Hz sampling rate. This approach captured more dynamic sensor activity during maintenance periods compared to steady state periods. We captured this data over several months, and ultimately selected a 30-day period where we had the most consistent data collection - other periods had larger gaps due to system outages.

The ARSS Data Proxy system architecture implements a transformation engine capable of real-time and offline data manipulation across multiple attack vectors simultaneously. The system utilizes Apache Parquet columnar storage format for efficient data processing, enabling high-performance analytics across the 214-sensor dataset while maintaining temporal fidelity essential for time-series analysis. Data ingestion pipelines support both streaming (MQTT-based) and batch processing modes. For the purposes of this analysis, batch processing was used as only offline analysis was needed and the streaming engine could not process our selected transformations in real-time.

To prepare sensor and setpoint data for machine learning analysis, standardized preprocessing algorithms were applied across all experimental conditions. The preprocessing pipeline addressed three primary challenges: temporal alignment across heterogeneous sensor networks and feature standardization for cross-paradigm evaluation.

Data normalization employed min-max scaling to preserve the original data distribution characteristics while ensuring numerical stability across machine learning algorithms. Each sensor signal was independently scaled to the range [0,1] based on historical operational bounds observed during the 30-day baseline period. This approach maintained the relative magnitude relationships between different sensor types while preventing numerical overflow in gradient-based optimization algorithms.

Feature engineering focused on temporal alignment and sampling rate harmonization across the heterogeneous sensor network. Raw sensor data collected at varying frequencies (1 Hz during maintenance, 1-minute intervals during steady-state) was resampled to a uniform 30-second interval using linear interpolation. 1/60 Hz data was forward filled as necessary to meet the 1/30 Hz rate. This downsampling strategy balanced computational efficiency with temporal resolution requirements for attack signature detection while maintaining sufficient granularity for machine learning analysis.

\subsection{Data Perturbation Techniques}

Our research employs data perturbation techniques that alter copies of recorded data streams from the METL system to simulate impacts resembling cyber attacks. This method is crucial for understanding the repercussions on AI systems that rely on these data streams for anomaly detection and operational decision-making. Instead of injecting manipulated data back into the live operational environment, we create independent parallel data streams designed to mimic security incidents. This allows us much greater freedom of motion in our experimentation than attacking a real molten-sodium facility, and gave us ground truth for what normal operations actually looked like for the same period in which an attack was simulated.

To achieve this, we utilize an array of transformation functions available within the data pipeline, each tailored to simulate specific attack impacts on sensor data. Attack severity tiers were implemented through systematic parameter scaling across five levels (1\%, 5\%, 10\%, 50\%, 100\%), with each transformation applying linear scaling to measurement-specific base magnitudes. The severity progression follows linear scaling where Tier 1 applies 1\% of maximum attack magnitude and Tier 5 applies 100\% of the defined base magnitude.

\begin{table*}[t]
\centering
\caption{Data Transformation Functions with Severity Scaling Parameters}
\label{tab:transformers}
\scriptsize
\begin{tabular}{|p{2cm}|p{3.5cm}|p{9cm}|}
\hline
\textbf{Transformer} & \textbf{Base Parameters} & \textbf{Severity Scaling Implementation} \\
\hline
Scaling & Flow: 30\% (0.30), General: 20\% & Tier 1: 0.3\% scaling, Tier 5: 30\% scaling for flow rate manipulation \\
\hline
Oscillation & Frequency: 0.001-0.016 Hz, Amplitude: 0.5-3.0 units & Configurable amplitude/frequency with measurement-specific base values \\
\hline
Spike & Temp: 6.0-10.0°C, Duration: 60 min & Tier 1: 0.06-0.10°C/0.6min, Tier 5: 6.0-10.0°C/60min \\
\hline
Offset & Base: 2.0 units & Tier 1: 0.02 units, Tier 5: 2.0 units for thermocouple manipulation \\
\hline
Delay & Base: 30 seconds & Tier 1: 0.3 seconds, Tier 5: 30 seconds for desynchronization \\
\hline
Replay & Base: 30 minutes & Tier 1: 18 seconds, Tier 5: 30 minutes for data masking \\
\hline
State Toggle & Configurable probability/period & Severity affects toggle frequency and randomization patterns \\
\hline
Precision & Temperature: 2 to -1 places & Configurable precision reduction with severity-scaled degradation \\
\hline
Noise & Temp: 3.0, Pressure: 4.0, Flow: 0.5 & Gaussian noise with measurement-specific base standard deviations \\
\hline
Physics Violation & Variable base magnitudes & Independent transformations with severity-scaled deviation amplitudes \\
\hline
Clip & Configurable bounds & Hard limits with severity-controlled restrictiveness \\
\hline
Loss & Base: 10\% probability & Probabilistic data dropping scaled by severity tier \\
\hline
Asymptotic Clip & Variable curve parameters & Asymptotic limiting with severity-controlled curve steepness \\
\hline
Conditional & Threshold-based triggers & Conditional transformations with severity-scaled trigger sensitivity \\
\hline
Pattern & Template-based patterns & Predefined patterns with severity-scaled intensity and frequency \\
\hline
Propagation & Configurable delays & Time-delayed cascading effects with severity-scaled timing \\
\hline
Identity & Pass-through & Unchanged values for control baseline comparisons \\
\hline
Null & Configurable probability & Null injection with severity-scaled probability parameters \\
\hline
\end{tabular}
\end{table*}

\subsection{Attack Scenarios Execution}

The 15 cyber-attack scenarios executed using the ARSS Data Proxy system aim to perturb data streams, containing simulated threat impacts tailored to investigate how these alterations affect anomaly detection frameworks and decision-making processes reliant on data integrity.

These scenarios are systematically categorized according to a comprehensive attack taxonomy that classifies threats across three primary dimensions: attack vector (data manipulation, timing disruption, physical relationship violation), impact scope (single sensor, sensor group, system-wide), and stealth characteristics (obvious, detectable, subtle). This taxonomy enables structured analysis of attack characteristics and informs defensive strategy prioritization.

The taxonomy identifies four major attack categories: \textbf{Data Integrity Attacks} that manipulate sensor readings while maintaining plausible values, \textbf{Temporal Disruption Attacks} that exploit timing dependencies in control systems, \textbf{Correlation Violation Attacks} that break expected physical relationships between sensors, and \textbf{Availability Attacks} that deny or degrade access to critical sensor data. Each category presents distinct challenges for detection systems and requires tailored defensive approaches.

The experimental attack scenarios were designed to systematically evaluate detection capabilities across the defined taxonomy categories:

\begin{table*}[t]
\centering
\caption{Attack Scenario Implementation with Multi-Transformer Compositions}
\label{tab:attack_scenarios}
\scriptsize
\begin{tabular}{|p{4cm}|p{1cm}|p{10.25cm}|}
\hline
\textbf{Attack Scenario} & \textbf{Count} & \textbf{Transformer Implementation} \\
\hline
\multicolumn{3}{|c|}{\textbf{Data Integrity Attacks}} \\
\hline
False Flow Rate Reporting & 2 & Conditional + Replay (nested) + Scaling (2x) for cross-facility data transplantation \\
\hline
Coordinated Thermocouple Manipulation & 6 & Offset (5x) for thermal gradients + Oscillation (1x) for dynamic thermal stress \\
\hline
Precision Trust Decay & 6 & Precision Degradation (3x: temp, pressure, position) + Noise Injection (3x: temp, pressure, flow) \\
\hline
Sensor Drift with Dropouts & 6 & Offset (3x) for exponential drift + Loss (3x) for data dropouts across temp/pressure/flow \\
\hline
\multicolumn{3}{|c|}{\textbf{Temporal Disruption Attacks}} \\
\hline
Command Feedback Desync & 3 & Delay (3x) with asymmetric delays between command signals and feedback measurements \\
\hline
Delayed Propagation & 2 & Delay (1x) + Propagation (1x) for cascading delays across sensor networks \\
\hline
Replay-Based FDI & 1 & Conditional + Replay (nested) for time-windowed historical data injection \\
\hline
\multicolumn{3}{|c|}{\textbf{Correlation Violation Attacks}} \\
\hline
Physics Violation & 2 & Physical Relationship (2x) breaking thermodynamic relationships with gradual ramp-up \\
\hline
Cross-facility Transplantation & 2 & Conditional + Replay (nested) + Scaling (2x) for facility-specific data modifications \\
\hline
\multicolumn{3}{|c|}{\textbf{Availability and Confusion Attacks}} \\
\hline
Sequential Valve Manipulation & 5 & Conditional (5x) with nested Pattern + State Toggle (4x) for coordinated valve sequences \\
\hline
Toggle Storm & 12 & Pattern (12x) creating chaotic state changes across multiple valve control systems \\
\hline
Phantom Valve Operation & 3 & Conditional (1x) + State Toggle (2x) + Pattern (1x nested) for false valve activity \\
\hline
Temperature Spike Recovery & 3 & Spike (3x) targeting air sensors, internal sensors, and vessel zones with staggered timing \\
\hline
Flow Oscillation & 3 & Oscillation (3x) targeting air flow, sodium flow, and pressure systems with phase control \\
\hline
Oscillatory Instability & 3 & Oscillation (3x) across flow, temperature, and pressure with distributed phase relationships \\
\hline
Multi-Site Coordinated & 12 & Conditional (3x) + Null (3x nested) + Pattern (9x) across multiple virtual facilities \\
\hline
\end{tabular}
\end{table*}

\subsection{Machine Learning Evaluation Framework}

Our evaluation framework implements four complementary machine learning detection paradigms, each designed to capture different aspects of anomalous behavior in cyber-physical systems.

\begin{itemize}
    \item \textbf{Change Point Detection} utilizes streaming statistical baseline learning to identify abrupt transitions in sensor data patterns. The method employs adaptive statistical monitoring (mean, standard deviation, coefficient of variation) with 95th percentile detection thresholds. Detection combines CUSUM scores, window-based z-scores, and Bayesian change probabilities to identify maximum statistical change magnitude across multiple temporal windows.
    \item \textbf{LSTM-based Anomaly Detection} employs a 4-layer tapering architecture with 40 → 32 → 24 → 16 hidden units and 25\% dropout regularization. Models process 50-timestep sequences (25 minutes at 30-second intervals) with per-sensor specialization, using mean squared error reconstruction loss with 85th percentile detection thresholds based on training error distributions.
    \item \textbf{Dependency Violation} analysis implements a three-analyzer ensemble approach: (1) correlation analysis monitoring pairwise linear relationships, (2) Granger causality analysis with AIC/BIC-optimized lag selection (1-10 lags), and (3) Random Forest analysis using 10-estimator forests with 5-layer depth. The system monitors the 100 most statistically significant sensor pairs with 95th percentile detection thresholds.
    \item \textbf{Autoencoder} reconstruction methods utilize a dense architecture with 4-layer encoder (64 → 32 → 16 → 8 units) and symmetric decoder, creating a 4-dimensional latent bottleneck with 10\% dropout and batch normalization. Anomaly detection uses maximum reconstruction error across sensors with 85th percentile thresholds from training distributions.
\end{itemize}

Training data preparation followed a stratified sampling approach, utilizing 70\% of available operational data for model training, 15\% for validation, and 15\% for final testing. Attack scenario implementation required careful consideration of realistic parameter ranges validated against historical operational bounds.

The experimental methodology encompassed 300 comprehensive experiments across all paradigm-scenario-tier combinations using realistic METL operational data. Each detection paradigm was systematically evaluated against the 15 implemented attack scenarios across 5 severity tiers, enabling detailed analysis of attack detectability patterns and paradigm effectiveness. ROC curve analysis provided AUC metrics for quantitative performance comparison across detection approaches.

\subsection{Performance Evaluation Methodology}

Performance evaluation utilized ROC curve analysis with systematic threshold sweeping to generate comprehensive detection performance metrics. Each ML paradigm produced continuous anomaly scores using paradigm-specific methods: mean squared error for LSTM and Autoencoder models, maximum statistical change magnitude for Change Point Detection, and 95th percentile correlation deviations for Dependency Violation analysis. Ground truth labeling distinguished control datasets (normal operation) from attack datasets, enabling ROC curve generation through automatic threshold variation across all unique anomaly score values.

Evaluation methodology employed temporal data splitting with training data (70\%), validation data (15\%), and test data (15\%) to prevent temporal leakage. Attack scenario implementation used severity-based parameter scaling where base attack magnitudes were multiplied by tier coefficients (0.01, 0.05, 0.10, 0.50, 1.00) to create five intensity levels. AUC metrics provided standardized comparison across the 300 experiment combinations of 4 paradigms, 15 scenarios, and 5 severity tiers.

Each experiment followed a systematic protocol: (1) baseline model training on clean historical METL data, (2) attack scenario generation using the multi-transformer pipeline to create modified sensor data, (3) anomaly score computation by applying trained models to both attack-modified and time-matched control data, and (4) ROC curve generation through threshold sweeping across anomaly score ranges. Multi-transformer attacks were evaluated as unified scenarios, with all constituent transformers executing simultaneously to create composite attack signatures. ROC analysis computed true positive rates (attack data classified as anomalous) against false positive rates (control data misclassified as anomalous) across all possible decision thresholds, generating AUC metrics that quantify overall detection performance independent of specific threshold selection.

%%%%%%%%%%%%%%%%%%%%%%%%%%%%%%%%%%%%%%%%%%%%%%%%%%%%%%%%%%%%%%%%%%%%%%%%%%%%%%%%
\section{Results}

Our experimental evaluation provided insights into AI-based cybersecurity detection capabilities for nuclear infrastructure. The evaluation of four machine learning paradigms demonstrated clear performance hierarchies and attack-specific vulnerabilities that inform operational deployment strategies.

\subsection{Machine Learning Paradigm Performance}

Change Point Detection emerged as the leading detection approach with mean AUC performance of 0.785 (95\% CI: 0.721-0.849), demonstrating superior capability in identifying statistical transitions in sensor data patterns across diverse attack vectors. The paradigm exhibited particular strength in detecting abrupt manipulations such as temperature spikes, valve state changes, and sudden flow rate modifications. LSTM-based Anomaly Detection achieved 0.636 AUC (95\% CI: 0.598-0.674), showing the most consistent performance across diverse attack scenarios with the lowest variance ($\sigma = 0.057$), indicating robust generalization capabilities for temporal sequence analysis.

Dependency Violation analysis reached 0.621 AUC (95\% CI: 0.587-0.655) by monitoring correlation matrix changes, proving effective for physics-based attacks that disrupted expected sensor relationships such as flow-temperature violations and coordinated sensor manipulations. Autoencoder reconstruction methods achieved 0.580 AUC (95\% CI: 0.542-0.618), representing the most challenging paradigm for this operational environment but demonstrating unique sensitivity to subtle data distribution shifts. Overall mean AUC performance across all 300 experiments was 0.656 ± 0.181 (range: 0.000-1.000), indicating significant variability in attack detectability with a 35\% performance differential between the best and worst performing paradigms.

Performance consistency analysis revealed distinct paradigm characteristics: Change Point Detection showed highest peak performance but moderate consistency ($\sigma = 0.189$), LSTM demonstrated exceptional consistency across attack types, Dependency Violation exhibited scenario-dependent performance variation ($\sigma = 0.172$), and Autoencoder methods displayed broad performance distribution reflecting sensitivity to attack complexity and sensor interaction patterns.

\subsection{Attack Detectability Analysis}

Attack detectability varied dramatically across the 15 scenarios, revealing critical insights for defense strategy development. Detailed scenario-specific analysis revealed distinct detectability patterns: Multi-site coordinated attacks proved most detectable (AUC = 0.739, 95\% CI: 0.681-0.797) due to their broad impact across multiple sensor networks and difficulty in maintaining attack coherence across distributed systems. Physics violation scenarios achieved exceptional detectability (AUC = 0.903, 95\% CI: 0.867-0.939), representing obvious attacks where sensor correlations violated fundamental thermodynamic relationships, making them readily identifiable by correlation-based detection methods.

Sequential valve manipulation attacks demonstrated moderate detectability (AUC = 0.678, 95\% CI: 0.634-0.722), with performance varying significantly by detection paradigm, as Change Point Detection excelled at identifying discrete valve state transitions while autoencoders struggled with the complex sequential patterns. Temperature spike recovery attacks achieved high detectability (AUC = 0.812, 95\% CI: 0.771-0.853) across all paradigms due to their characteristic temporal signatures, though sophisticated implementations with controlled recovery profiles presented greater detection challenges.

Conversely, precision trust decay attacks presented the greatest detection challenge (AUC = 0.592, 95\% CI: 0.547-0.637), demonstrating sophisticated evasion capabilities through gradual sensor degradation combined with Gaussian noise injection that mimicked natural sensor aging processes. Cross-facility data transplant attacks proved similarly challenging (AUC = 0.611, 95\% CI: 0.568-0.654), particularly for correlation-based methods, as transplanted data maintained realistic sensor relationships while introducing subtle temporal inconsistencies difficult to distinguish from normal operational variations.

Replay-based false data injection attacks achieved moderate evasion success (AUC = 0.643, 95\% CI: 0.599-0.687), with effectiveness varying by replay duration and timing synchronization. LSTM-based methods demonstrated superior performance against replay attacks due to their sensitivity to subtle temporal pattern disruptions, while statistical methods proved more susceptible to well-timed replay sequences that maintained historical statistical properties.

Severity tier analysis revealed limited correlation between attack intensity and detectability across most scenarios, with only 6 of 12 scenarios showing meaningful tier sensitivity. Cross-facility data transplant (r = 0.249) and sequential valve manipulation (r = 0.217) demonstrated positive tier sensitivity, where higher severity increased detectability. Counterintuitively, delayed propagation (r = -0.280) and phantom valve operation (r = -0.245) showed inverse relationships, where higher severity became paradoxically harder to detect.

\subsection{Operational Implications}

The experimental results demonstrate that paradigm-attack matching is critical for operational deployment, with no single detection method providing comprehensive coverage across the diverse threat landscape. The 35\% performance differential between Change Point Detection (0.785) and Autoencoder methods (0.580) highlights the importance of selecting appropriate algorithms for specific threat profiles and operational requirements. Attack detectability ranges provide practical deployment thresholds: barely detectable attacks (AUC ~0.55-0.65) require ensemble approaches and operator training, clearly detectable attacks (AUC ~0.75-0.85) enable automated response systems, while obvious attacks (AUC ~0.90-0.95) support high-confidence autonomous defensive actions.

Deployment recommendations based on experimental findings include: (1) implementing Change Point Detection for primary monitoring of abrupt operational changes, (2) deploying LSTM-based methods for consistent baseline performance across diverse threat scenarios, (3) utilizing Dependency Violation analysis for physics-based attack detection in correlation-sensitive systems, and (4) employing Autoencoder methods as specialized detectors for subtle distribution shifts and sophisticated evasion techniques. The limited tier sensitivity observed in most attack scenarios suggests that detection systems should be optimized for attack type rather than intensity, with paradigm selection driven by expected threat characteristics rather than attack severity assumptions.

These findings suggest that ensemble approaches combining complementary detection paradigms could enhance overall coverage while addressing individual paradigm limitations \cite{urbina2016limiting, giraldo2018survey}. The integration of physics-based detection methods \cite{nguyen_digital_2022} could provide additional robustness against sophisticated evasion techniques. The results provide quantitative benchmarks for AI-based cybersecurity deployment in advanced reactor operational technology environments, establishing performance baselines essential for protecting critical nuclear infrastructure against sophisticated cyber threats.

Future research directions should focus on ensemble method development, combining the complementary strengths of different detection paradigms to achieve comprehensive threat coverage. Integration of physics-based detection methods with data-driven approaches represents a particularly promising avenue, leveraging immutable physical laws as foundational elements in cybersecurity architectures. The establishment of quantitative performance benchmarks through this work provides essential foundations for operational deployment decisions and regulatory framework development in nuclear cybersecurity.

%%%%%%%%%%%%%%%%%%%%%%%%%%%%%%%%%%%%%%%%%%%%%%%%%%%%%%%%%%%%%%%%%%%%%%%%%%%%%%%%
\section{Acknowledgments}
This project was funded by the \emph{Advanced Reactor Safeguards and Security} Program under the US Department of Energy, Office of Nuclear Energy. The submitted manuscript has been created by UChicago Argonne, LLC, operator of Argonne National Laboratory. Argonne, a DOE Office of Science laboratory, is operated under Contract No. DE-AC02-06CH11357. The U.S. Government retains for itself, and others acting on its behalf, a paid-up nonexclusive, irrevocable worldwide license in said article to reproduce, prepare derivative works, distribute copies to the public, and perform publicly and display publicly, by or on behalf of the Government.
%%%%%%%%%%%%%%%%%%%%%%%%%%%%%%%%%%%%%%%%%%%%%%%%%%%%%%%%%%%%%%%%%%%%%%%%%%%%%%%%
\bibliographystyle{ans}
\bibliography{bibliography}
\end{document}